\documentclass[lettersize,journal]{IEEEtran}
\usepackage{amsmath,amsfonts}
\usepackage{array}
\usepackage[caption=false,font=normalsize,labelfont=sf,textfont=sf]{subfig}
\usepackage{textcomp}
\usepackage{stfloats}
\usepackage{url}
\usepackage{verbatim}
\usepackage{graphicx}
\usepackage{capt-of}
\def\BibTeX{{\rm B\kern-.05em{\sc i\kern-.025em b}\kern-.08em
    T\kern-.1667em\lower.7ex\hbox{E}\kern-.125emX}}
\usepackage{balance}
\usepackage{algorithm}
\usepackage{algpseudocode}
\usepackage{enumitem}
\usepackage{ragged2e}
\usepackage{placeins}
\usepackage[hidelinks]{hyperref}

\begin{document}

\title{%
  {\fontsize{18}{22}\selectfont
   Micro-Splatting: Multistage Isotropy-informed Covariance Regularization Optimization for High-Fidelity 3D Gaussian Splatting }%
}

\author{%
  Jee Won Lee\textsuperscript{1,2}, 
  Hansol Lim\textsuperscript{1,2}, 
  Sooyeun Yang\textsuperscript{1,2}, 
  and Jongseong Brad Choi\textsuperscript{1,2*}\\[0.3ex]
  {\footnotesize\normalfont%
    \textsuperscript{1} Department of Mechanical Engineering, State University of New York, Korea, Incheon, South Korea\\[0.3ex]
    \textsuperscript{2} Department of Mechanical Engineering, State University of New York, Stony Brook, NY, United States\\[0.3ex]
    \textsuperscript{*}Corresponding author
  }%

\begin{center}
  \includegraphics[width=0.75\textwidth]{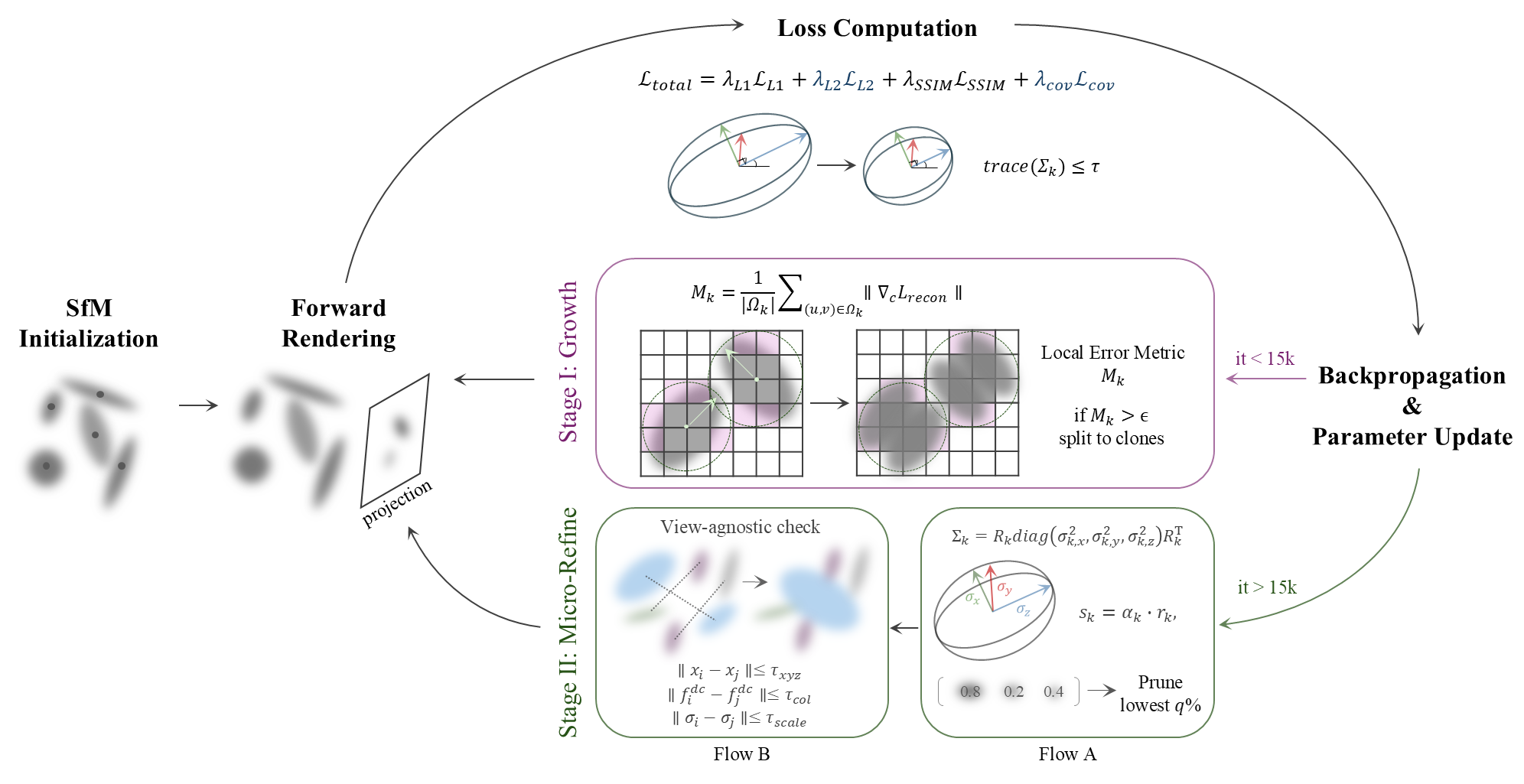}\\[0.5ex]
  {\footnotesize
   \begin{minipage}{\textwidth}
     \raggedright
     Fig. 1. Overall Training Pipeline of Micro-splatting. This diagram shows how our method progresses from SfM points to fully optimized 3D Gaussians. Starting from the SfM points, we render the scene, compute a mixed photometric loss($L_{1}$ + $L_{2}$ + SSIM) with a trace-based covariance penalty, and update parameters. Stage I (Growth, 15k iters) split only those Gaussians whose view-space footprint shows high local reconstruction error. Stage II (Micro-Refine, 15 > iters) prunes low-importance splats and merges redundant neighbors via view-agnostic checks. The end-to-end loop needs no auxiliary networks or post-processing and yields compact, near-isotropic, detail-preserving splats.  
   \end{minipage}
  }
\end{center}

\thanks{\noindent This work was supported by the National Research Foundation of Korea (NRF) grant funded by the Korea government (MSIT) (Grant No.\ RS-2022-NR067080 and RS-2025-05515607).}}

\maketitle

\begin{abstract}
High‐fidelity 3D Gaussian Splatting methods excel at capturing fine textures but often overlook model compactness, resulting in massive splat counts, bloated memory, long training, and complex post-processing. We present Micro-Splatting: Multistage Isotropy-informed Covariance Regularization Optimization for High-Fidelity 3D Gaussian Splatting, a unified pipeline that preserves visual detail while drastically reducing model complexity without any post-processing or auxiliary neural modules. In Stage I (Growth), we introduce a trace-based covariance regularization to maintain near-isotropic Gaussians, mitigating low-pass filtering in high-frequency regions and improving spherical-harmonic color fitting. We then apply gradient-guided adaptive densification that subdivides splats only in visually complex regions, leaving smooth areas sparse. In Stage II (Refinement), we prune low-impact splats using a simple opacity–scale importance score and merge redundant neighbors via lightweight spatial and feature thresholds, producing a lean yet detail-rich model. On four object-centric benchmarks, Micro-Splatting reduces splat count and model size by up to 60\% and shortens training by 20\%, while matching or surpassing state-of-the-art PSNR, SSIM, and LPIPS in real-time rendering. These results demonstrate that Micro-Splatting delivers both compactness and high fidelity in a single, efficient, end-to-end framework.   
\end{abstract}
\section{Introduction}
\IEEEPARstart High-fidelity 3D scene representation is a cornerstone of modern computer vision and graphics, enabling applications from virtual reality to autonomous navigation. Traditional explicit models such as voxels, meshes, and point clouds often suffer from discontinuities and view-dependent artifacts ~\cite{ref1}, \cite{ref2}. Neural Radiance Fields (NeRF) introduced a continuous volumetric formulation that produces photorealistic novel-view rendering but incurs prohibitive training and inference costs due to its reliance on deep MLPs ~\cite{ref3},\cite{ref4}. 
\\\indent Recently, 3D Gaussian Splatting (3DGS) has emerged as a powerful alternative, representing scenes as collections of oriented Gaussian primitives that can be rasterized in real time ~\cite{ref5},\cite{ref6}. By learning per-splat position, covariance, opacity, and spherical-harmonic color coefficients, 3DGS achieves much faster rendering than NeRF with competitive fidelity. However, capturing fine textures and sharp edges typically demands either oversized or highly anisotropic covariance radii, which act as directional low-pass filters and blur fine details, or uniformly tiny splats, which cause an explosion in point count and memory usage. This trade-off is further compounded by anisotropy biasing spherical-harmonic color fitting, motivating our introduction of a trace-based covariance regularization to maintain near-isotropic support while adaptively allocating splats where detail is most needed.
\\\indent Recent works push 3DGS toward finer detail, often at substantial cost ~\cite{ref7}. AbsGS ~\cite{ref8} adaptively splits splats in high-frequency regions using gradient heuristics; FreGS ~\cite{ref9} adds a global frequency regularizer that drives high-frequency content throughout training; Pixel-GS ~\cite{ref10} aligns splats to pixel footprints to fight overlap and under-coverage. These strategies improve sharpness, but they commonly inflate point counts requiring millions of splats to represent the scene, increase memory and training time, and, in some cases, add multi-stage or auxiliary neural components (e.g., Taming 3DGS ~\cite{ref11}) that further raise complexity.  
\\\indent First, we incorporate a covariance regularization term that enforces isotropic constraints on each gaussian splat. Isotropic kernels apply identical spatial extent in all directions, ensuring uniform treatment of high-frequency content without biasing along any axis. In contrast, anisotropic kernels can over-smooth details along their major axis, leading to uneven loss of fine structure ~\cite{ref7},\cite{ref8}. This uniform behavior aligns naturally with spherical harmonic-based color representations, which assume rotational symmetry, and further improves view-dependent color fidelity. By penalizing large or elongated covariances, our regularization keeps each splat compact and nearly spherical, preserving localized high-frequency details. 
\\\indent A second thread targets compactness but risks oversimplification. EDC ~\cite{ref12} performs learned compaction and distillation to shrink models, and Mini-Splatting 2 ~\cite{ref13} uses hierarchical subdivision and consolidation to reduce footprint. While these approaches can be lightweight at inference, they tend to bias capacity toward object-centric regions and blur or wash out distance background detail.
\\\indent In this paper, we propose Micro-Splatting, a two-stage, fully in-training pipeline that addresses the long-standing trade-off between fine-detail fidelity and model compactness in 3D Gaussian Splatting. Our approach integrates detail recovery and model reduction into a single optimization loop, without any post-processing or auxiliary neural modules. The core contributions are:\vspace{-0.3em}
\begin{itemize}
 \item \textbf{Trace-based isotropy in Stage~I.} We penalize oversized or highly anisotropic Gaussians via a simple trace and scale cap, keeping splats near-isotropic. This both mitigates the Gaussian low-pass bias, preserving high-frequency detail, and reduces directional bias in SH color fitting. Crucially, exceeding the cap also triggers splitting, so the constraint directly drives where capacity is added.
 \item \textbf{Gradient-guided densification in Stage~I.} After each update we compute a local reconstruction-gradient score over each splat’s view-space footprint and adapt the split threshold. Only splats covering high-gradient, high-frequency regions subdivide where flat regions remain sparse. This targets capacity where it matters and avoids blanket growth.
 \item \textbf{Two-stage refinement module in Stage~II.} We first prune low-impact splats using a cheap, interpretable importance score (\(\mathrm{opacity}\times \mathrm{max}\ {scale}\)). We then merge redundant neighbors with simple, view-agnostic checks. Together these steps remove the long tail and collapse duplicates, yielding a lean yet detail-rich model.
 \item \textbf{Simple computation.} Unlike methods that add auxiliary networks or post-training passes, our sequence uses only lightweight computations within a single training loop.
\end{itemize}

\indent By integrating these techniques, Micro-Splatting achieves a simple yet powerful mechanism that deliver a robust, scalable solution for compact, high-fidelity 3DGS model. Our balanced design minimizes severe trade-offs and matches or exceeds state-of-the-art metrics. 

\section{Related Works}
\subsection{Novel View Synthesis}
\indent  The field of Novel View Synthesis (NVS) has been transformed by Neural Radiance Fields (NeRF) ~\cite{ref4}, which encode scene as continuous volumetric functions via multilayer perceptrons, enabling photorealistic rendering from arbitrary viewpoints. Despite their visual quality, NeRF-based methods are hampered by lengthy training times and slowS inference. Addressing these limitations, EfficientNeRF ~\cite{ref14} introduces optimizations that reduce both training and rendering costs, yet they remain constrained by the inherent overhead of neural networks. A significant paradigm shift occurred with the emergence of 3D Gaussian primitives that achieve real-time rendering through fast rasterization through continuous kernels. 

\subsection{Detail-Preserving Enhancements in 3DGS }
\indent Building on 3DGS, recent works target the loss of fine details caused by large or anisotropic splats. AbsGS ~\cite{ref8} introduces a gradient-driven splitting rule that adaptively subdivides oversized Gaussians in high-frequency regions to sharpen local textures. FreGS ~\cite{ref9} adds a frequency-domain regularizer that progressively encourages higher-frequency content during training. Spec-Gaussian ~\cite{ref15} augments 3DGS with an anisotropic appearance model to better reproduce view-dependent effects such as specular highlights. Pixel-GS ~\cite{ref10} aligns primitives with pixel footprints to better match per-pixel residuals and improve fine-scale fidelity. TrimGS ~\cite{ref16} performs contribution-based trimming and scale-driven splitting, periodically removing low-contribution Gaussians and subdividing oversized ones to keep kernels small and recover high-frequency geometry.

\indent However, recent detail-enhancing methods generally prioritize fidelity over compactness as they introduce complex losses and rarely include a built-in consolidation step (pruning/merging) where splat counts and memory often grow unchecked. Their refinement cues are typically global or pixel-aligned rather than footprint-aware, which can oversample smooth regions and inflate training time. Our work targets these gaps by keeping the pipeline simple while actively controlling where and how much to densify and then compress.

\subsection{Isotropy and Shape Regularization in 3DGS }
\indent Complementary to the detail-preserving methods, several works refine 3DGS by encouraging the Gaussians to be isotropic. Prior isotropy efforts either enforce hard isotropic kernels for speed ~\cite{ref17} or penalize eigenvalue imbalance via effective-rank or condition-number objectives or shape-aware splitting ~\cite{ref18}, \cite{ref19}. In contrast, we apply a size-aware trace cap with a hinge penalty. This keeps splats compact and near-isotropic, without forbidding beneficial anisotropy. Crucially, violations trigger splitting, directly coupling the constraint to topology changes so capacity is added exactly where the constraint would other be breached. Our mechanism is analytic and in-training and alongside a footprint-local gradient trigger, which together yield compact yet detail-preserving models.

\subsection{Compact Representations for 3DGS }
\indent In parallel, a body of work has explored compressing 3DGS models to reduce storage and computational demands. Taming 3DGS ~\cite{ref11} applies visibility- and contribution-based pruning after training, removing low-impact splats to shrink model size. Mini-Splatting 2 ~\cite{ref13} adopt quantization and codebook-based encodings to store Gaussian parameters compactly, while EDC ~\cite{ref12} uses an importance-driven decimation and clustering algorithm to merge spatially and spectrally similar primitives. Several other methods extend this trend by integrating more sophisticated compression strategies, including LightGaussian ~\cite{ref20}, which incorporates Gaussian volume and opacity into an importance metric followed by quantization for storage savings, and High-Fold Pruning ~\cite{ref21}, which uses opacity-assisted ranking to aggressively prune redundant splats. TrimGS~\cite{ref16} periodically prunes low-contribution splats and enforces small kernels via scale-driven splitting, yielding leaner models while preserving geometric detail.
\\\indent More recent approaches employ neural networks or heavy attribute compression frameworks to achieve higher compression ratios. Scaffold-GS ~\cite{ref22} encodes Gaussian attributes via anchor-based features refined by per-view MLPs, achieving strong compression at the cost of additional inference passes. LocoGS ~\cite{ref23} represents nearly all Gaussian attributes through a learned neural field, reducing storage but adding per-query neural evaluations. Optimized Minimal 3DGS (OMG) ~\cite{ref24} integrates a lightweight neural field with sub-vector quantization to balance irregularity and continuity in sparse Gaussian sets. Compressed 3DGS ~\cite{ref25} adopts residual vector quantization with entropy coding, while Reducing the Memory Footprint of 3DGS ~\cite{ref26} applies hierarchical neural quantization for extreme size reduction. Although these techniques can produce highly compact models, they typically operate as post-training stages, rely on additional neural components that increase implementation complexity and inference cost, or optimize for either compactness or fidelity in isolation. In contrast, our method jointly optimizes both within the same training loop using only lightweight analytic operations, eliminating the need for post-processing or auxiliary neural modules.

\subsection{Visual and Photometric Effects of Splat Anisotropy}
\indent Gaussian splats inherently exhibit a low-pass filtering effect, smoothing out high-frequency details within their spatial support ~\cite{ref6}. When the covariance matrix is anisotropic, having significant differences in the eigenvalues, this smoothing becomes directional, blurring details preferentially along the elongated axis ~\cite{ref27}. This geometric imbalance not only degrades fine spatial structures but also biases spherical-harmonic color fitting ~\cite{ref28}. Fig. 2 (Top) illustrates the geometric distinction where anisotropic splats exhibit stretched support that amplifies blur along one direction whereas isotropic splats maintain uniform detail preservation in all directions. 
\\\indent The photometric impact of this anisotropy is shown in Fig. 2 (Bottom), which compares the SH projections at degrees L=1, 2, 3 for isotropic and anisotropic splats against the ground truth. Anisotropic splats consistently produce higher root mean square error (RMSE) and introduce directional distortions, particularly in high-frequency components ~\cite{ref18}. These observations motivate our use of trace-based covariance regularization to maintain near-isotropic splats during training, mitigating both geometric and photometric degradation. 

\begin{figure}[!t]
\centering
  \includegraphics[width=\linewidth]{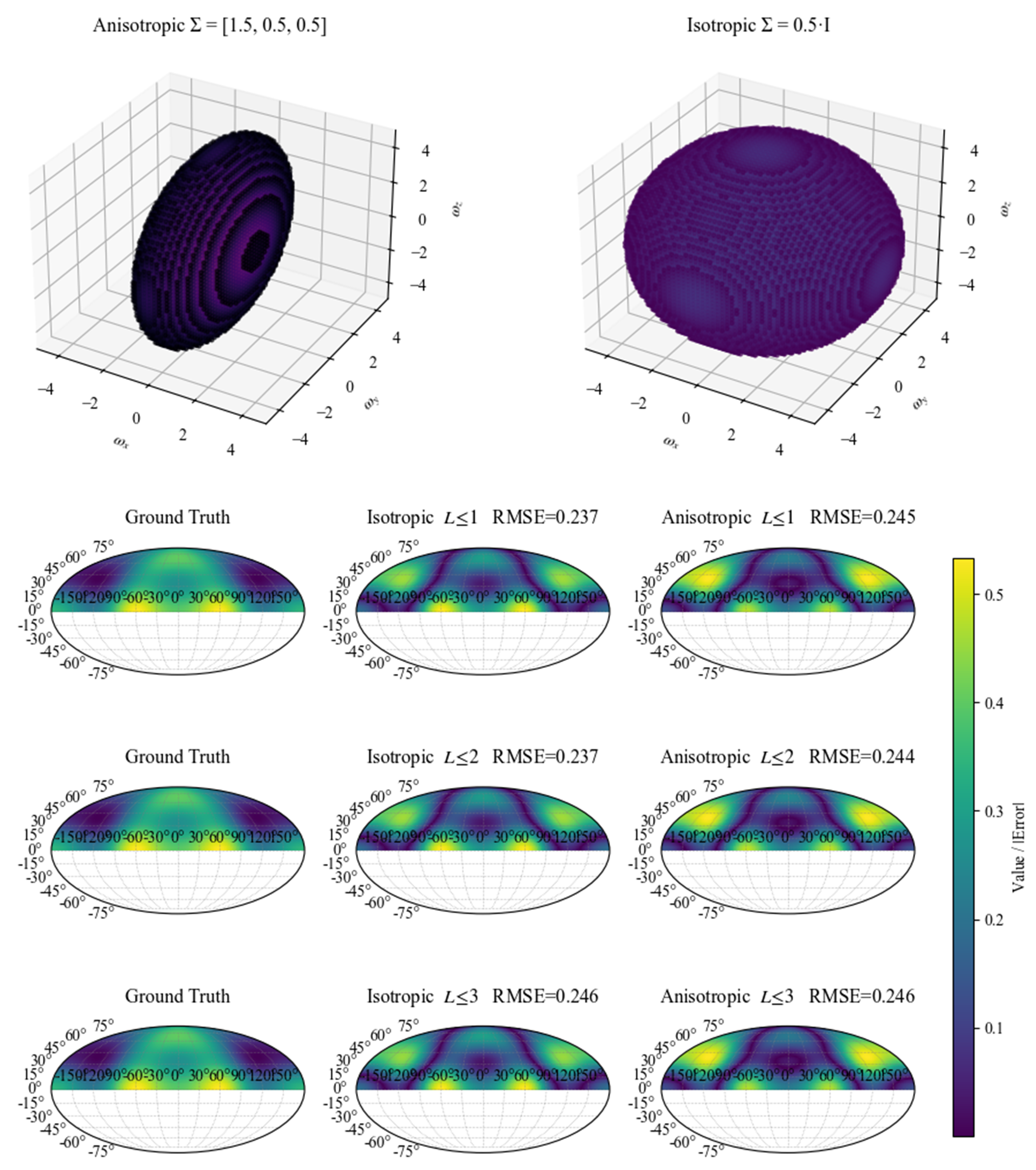}\\[2ex]
  {\footnotesize
   \parbox{\columnwidth}{
     Fig. 2. \textit{Visual and Photometric Effects of Splat Anisotropy.} {(Top)} Geometric representation of anisotropic and isotropic Gaussians. The anisotropic splat (\(\Sigma = [1.5, 0.5, 0.5]\)) exhibits elongated support along one axis, producing directional smoothing, whereas the isotropic splat (\(\Sigma = 0.5I\)) preserves uniform detail in all directions. {(Bottom)} Photometric impact on spherical-harmonic (SH) fitting for degrees \(L = 1, 2, 3\). Anisotropic splats introduce directional bias, leading to higher RMSE and visible distortions in high-frequency components compared to isotropic splats.
   }
  }
  \label{fig:anisotropy}
\end{figure}

\section{Preliminaries}
\subsection{3D Gaussian Splatting Fundamentals }
\indent We represent a scene as a set of  anisotropic 3D Gaussian primitives 

\begin{equation}
\label{deqn_ex1}
  G = \left\{ \left( \boldsymbol{\mu}_i, \Sigma_i, c_i, \alpha_i \right) \right\}_{i=1}^N ,
\end{equation}
where $\boldsymbol{\mu}_i \in \mathbb{R}^3$ is the mean position, 
$\Sigma_i \in \mathbb{R}^{3\times 3}$ is the covariance matrix defining spatial extent and orientation, $c_i$ encodes view-dependent color via spherical harmonics (SH), and $\alpha_i \in [0,1]$ denotes opacity. Each $\Sigma_i$ is positive-definite and can be decomposed as
\begin{equation}
\label{deqn_ex2}
  \Sigma_i = R_i S_i S_i^{T} R_i^{T},
\end{equation}
where $R_i \in \mathrm{SO}(3)$ is a rotation matrix and  $S_i \in \mathbb{R}^{3\times 3}$ is diagonal with entries proportional to the splat’s standard deviations along its principal axes.  Differences among these eigenvalues determine the splat’s anisotropy, which impacts both geometric sharpness and SH color fitting. 
\\\indent During rendering, each Gaussian is projected to the image plane, producing a 2D ellipse footprint from the transformed covariance 
\begin{equation}
\label{deqn_ex3}
  \Sigma_i' = J_i \Sigma_i J_i^{T},
\end{equation}
where $J_i$ is the Jacobian of the projection. The splats are rasterized in front-to-back order, blending their contributions according to alpha compositing. The spatial extent defined by $\Sigma_i$ governs the degree of local smoothing: large, elongated splats act as directional low-pass filters, while compact, near-isotropic splats preserve high-frequency details.

\subsection{L2-Enhanced Photometric Loss}
\indent The primary training objective in 3D Gaussian Splatting is to minimize the difference between rendered images and ground-truth views~\cite{ref6}. Let $I_{r}$ and $I_{t}$ denote the rendered and target images, respectively. The loss used in the original 3DGS formulation is a combination of an $L_{1}$ loss for structural accuracy and a perceptual similarity term such as SSIM:

\begin{equation}
\label{deqn_ex4}
L_{\mathrm{total}} = \lambda_{1} \left\lVert I_{r} - I_{t} \right\rVert_{1}  + \lambda_{\mathrm{SSIM}} \left( 1 - \mathrm{SSIM}\left( I_{r}, I_{t} \right) \right),
\end{equation}
This formulation encourages both pixel-wise fidelity and perceptual consistency across views. 
\\\indent In our work, we extend this objective with an additional  term, which further penalizes large deviations and helps stabilize optimization in later stages of training. The combined loss balances fine detail preservation with robustness to noise thereby forming the basis of our total training objective. 

\section{Micro-Splatting}
\subsection{Overview and Motivation}
\indent Micro-Splatting is designed to maintain the sharpness of 3D Gaussian Splatting reconstructions while keeping the model size, memory footprint, and training time low. Unlike approaches that rely on post-training compression or auxiliary neural modules, as illustrated in Fig. 1, Micro-Splatting is a single, in-training pipeline composed entirely of lightweight, analytic checks. 
\\\indent In Stage I (Growth), a trace-based covariance regularizer encourages splats to remain near-isotropic, reducing frequency-dependent blur. Growth is further guided by a footprint-local error metric, ensuring that new splats are added only in high-frequency regions where additional capacity is most beneficial. In Stage II (Micro-Refine), a simple importance score identifies and prunes low-impact splats, while redundant neighbors are merged using view-agnostic comparisons of position, color, and scale. This two-stage process yields a compact set of high-value splats that preserve fine detail while minimizing computational cost. 
\subsection{Covariance Regularization and Loss Optimization}
\noindent\textbf{Trace-based regularizer.} Micro-Splatting encourages each splat to remain compact and near-isotropic, preserving high-frequency detail and avoiding directional blur in SH color fitting. 

\begin{figure}[!t]
\centering
  \includegraphics[width=0.9\linewidth]{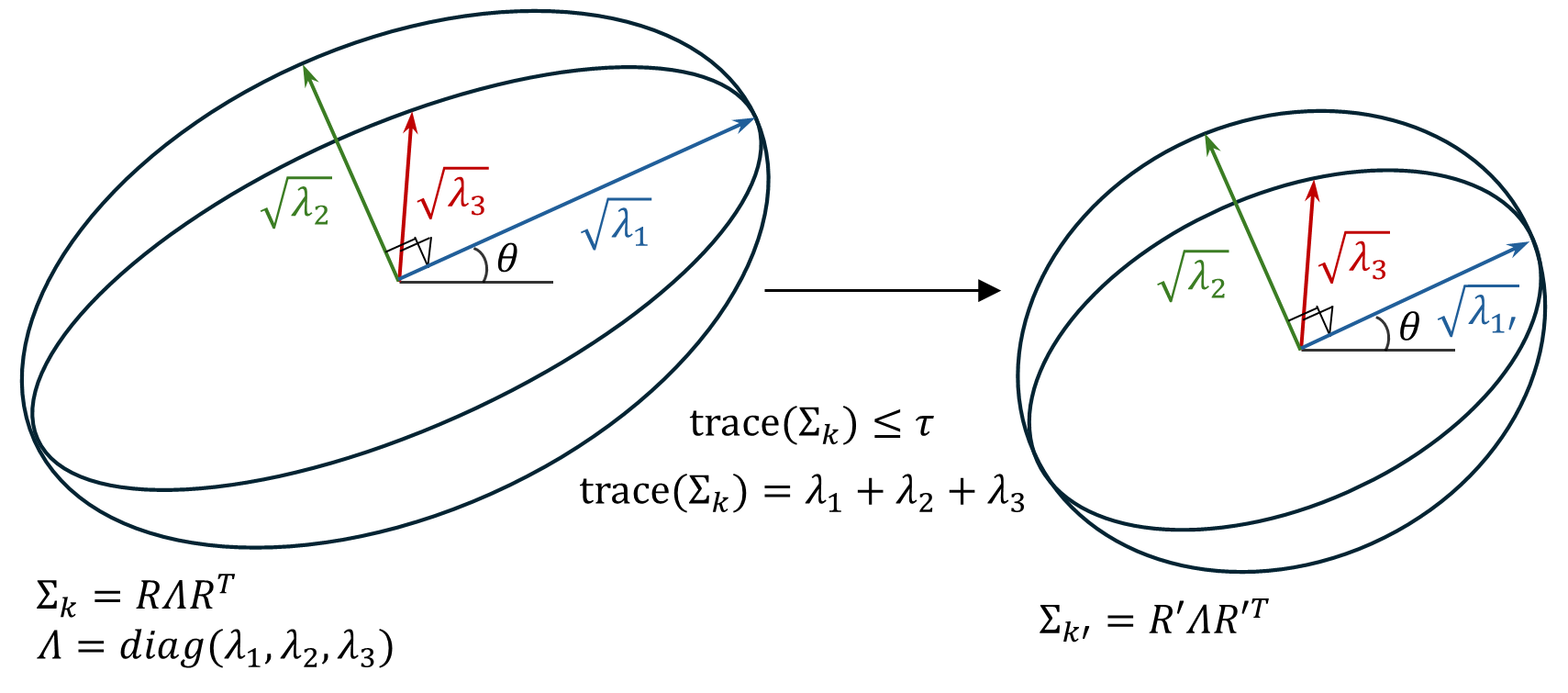}\\[2ex]
  {\footnotesize
   \parbox{\columnwidth}{
     Fig. 3. Eigen-Decomposition and Trace Constraint on the 3D Gaussian Covariance. 
This figure shows a Gaussian’s covariance $\Sigma_{k}$ decomposed into $R \Lambda R^{T}$, 
where $\lambda_{1}, \lambda_{2}, \lambda_{3}$ define the ellipsoid’s axes. 
By bounding $\mathrm{trace}(\Sigma_{k}) = \lambda_{1} + \lambda_{2} + \lambda_{3} \le \tau$, 
we limit how large these axes can grow, preserving near-isotropy and preventing excessive elongation.
   }
  }
\end{figure}

\indent Let $\Sigma_{k}$ denote the covariance of splat $k$ in eigen (axis-length) parameterization:
\begin{equation}
\Sigma_{k} = R_{k}\,\operatorname{diag}\!\left(\lambda_{k,1},\,\lambda_{k,2},\,\lambda_{k,3}\right) R_{k}^{T},
\end{equation}
where $R_{k} \in \mathrm{SO}(3)$ is a rotation matrix and $\lambda_{k,1}, \lambda_{k,2}, \lambda_{k,3} > 0$ are the squared axis lengths. The trace then becomes
\begin{equation}
\operatorname{tr}\!\left(\Sigma_{k}\right) = \lambda_{k,1} + \lambda_{k,2} + \lambda_{k,3}.
\end{equation}
We impose a trace cap $\tau$ and apply a hinge penalty $P_k$ per splat:
\begin{equation}
P_k = \max\!\big(\operatorname{tr}(\Sigma_{k}) - \tau,\, 0\big), \qquad
L_{\mathrm{cov}} = \sum_{k} P_k.
\end{equation}
Bounding the trace limits the overall spatial extent, preventing any axis from elongating excessively and thus maintaining near-isotropy.
\\\indent Unlike prior isotropy terms (e.g., Isotropic\mbox{-}GS~\cite{ref17}) that act as passive regularizers, 
here any splat with $P_{k}>0$ is also split, directly coupling the regularization with topology changes in the growth stage. 
This active linkage ensures that capacity is added only where the isotropy constraint would otherwise be violated (Fig.~3).
\par
\vspace{1\baselineskip}
\noindent\textbf{Photometric loss with L2 term} Building on the photometric reconstruction loss in Eq. 4, we incorporate an additional $L_{2}$ term provides error-proportional gradients that help correct subtle residuals. Specifically,
\begin{equation}
L_{2} = \frac{1}{N} \sum_{i=1}^{N} \left\lVert \hat{I}_{i} - I_{i} \right\rVert_{2}^{2},
\end{equation}
where $\hat{I}_{i}$ and $I_{i}$ denote the rendered and target pixel intensities, respectively.
Unlike $L_{1}$, whose gradient magnitude $\frac{\partial L_{1}}{\partial e}$ is constraint with respect to the pixel error$e_i$, the gradient of $L_{2}$ scales linearly as
$\frac{\partial L_{2}}{\partial e_i} = 2e_i$
This proportionality enables finer correction of small errors, improving convergence and enhancing edge fidelity in pixel-supervised tasks. Prior studies have shown that combining $L_{1}$ and $L_{2}$ can yield both stability and perceptual quality~\cite{ref28,ref29}.

\noindent Our total loss is therefore:
\begin{equation}
L_{\mathrm{total}} = \lambda_{1} L_{1} + \lambda_{2} L_{2} + \lambda_{\mathrm{SSIM}} L_{\mathrm{SSIM}} + \lambda_{\mathrm{cov}} L_{\mathrm{cov}},
\end{equation}
with $\lambda_{1}=\lambda_{2}=0.5$ unless otherwise noted. Together with the trace-based covariance constraint, this $L_{2}$-augmented photometric loss provides a unified training objective for Stage~I growth and Stage~II refinement, enabling compact, detail-preserving splats without the need for auxiliary networks or post-processing.

\begin{figure}[!t]
\centering
  \includegraphics[width=0.85\linewidth]{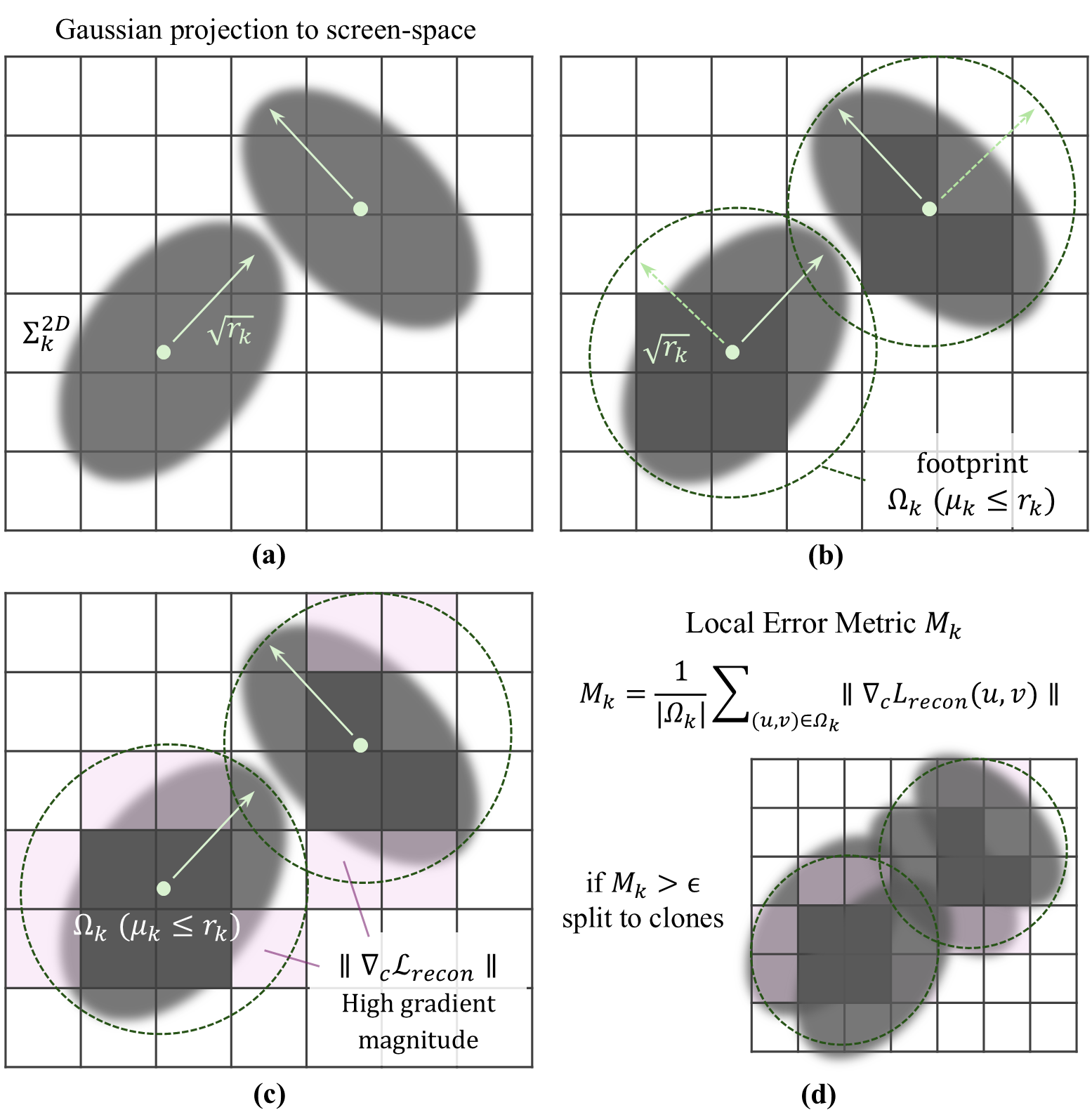}\\[2ex]
  {\footnotesize
   \parbox{\columnwidth}{
     Fig. 4. Stage I: Growth Phase. (a) Project each Gaussian and estimate its on-screen footprint. (b) Define a circular support around the project mean. (c) Measure reconstruction-loss gradients to detect under-fit, high-frequency regions. (d) Split into smaller clones to increase density only at details When the local score is high.
   }
  }
\end{figure}

\subsection{Two-Staged Refinement Module}\par
\vspace{1\baselineskip}
\noindent\textbf{Stage I. Growth Phase.} After each gradient step, we identify regions where the model still lacks detail and adds splats only in those locations (see Fig. 4):

\begin{description}[font=\itshape, labelsep=0.6em, leftmargin=2.6em, style=nextline, itemsep=0.3em]
  \item[(a)] Each Gaussian is projected into screen space, yielding its 2D covariance $\Sigma_{k}^{2\mathrm{D}}$.
  We define a footprint radius as:
  \begin{equation}
  r_{k} = \sqrt{\,\lambda_{\max}\!\left(\Sigma_{k}^{2\mathrm{D}}\right)}.
  \end{equation}
  where $\lambda_{\max}(\cdot)$ is the largest eigenvalue. This provides a constant-time proxy for the splat’s apparent size.

  \item[(b)] Around the projected mean $\boldsymbol{\mu}_{k}^{2\mathrm{D}}$, we define the circular footprint:
  \begin{equation}
  \Omega_{k} = \left\{ (u,v) : \left\lVert (u,v) - \boldsymbol{\mu}_{k}^{2\mathrm{D}} \right\rVert \le r_{k} \right\}.
  \end{equation}
  which approximates the pixel region influenced by the splat.

  \item[(c)] Using the same forward/backward pass as training, we read per-pixel reconstruction-loss gradients for $L_{\mathrm{recon}}$ in Eq.~9 and compute a magnitude map:
  \begin{equation}
  g(u,v) = \left\lVert \nabla_{c}\, L_{\mathrm{recon}}(u,v) \right\rVert .
  \tag{12}
  \end{equation}
  The local error score for a splat is the footprint average:
  \begin{equation}
  M_{k} = \frac{1}{\lvert \Omega_{k} \rvert} \sum_{(u,v)\in \Omega_{k}} g(u,v).
  \tag{13}
  \end{equation}
\indent We set $\varepsilon$ adaptively at each iteration as the $p$-th percentile of the current $\{M_k\}$ distribution, making the trigger scene-agnostic and robust.
  \item[(d)] If $M_{k}>\varepsilon$, we split $G_{k}$ into smaller clones by perturbing means inside $\Omega_{k}$ along the principal axes, halving scales, and copying color and opacity to avoid holes. In addition, if the trace penalty for $G_{k}$ is active ($P_{k}>0$, from Eq.~7), we split to prevent elongated kernels from persisting. 
\end{description}
\indent This process runs until iteration $T_{\mathrm{refine}}$ (15k in our experiments), concentrating many small, near-isotropic splats in high-frequency regions while leaving flat regions sparsely sampled.\par
\vspace{1\baselineskip}
\noindent\textbf{Stage II. Micro-Refine Phase} 
\\\indent After the growth phase, the splat set may still contain low-importance primitives. Stage II removes these inefficiencies through two lightweight, view-agnostic operations (see Fig. 5). 

\noindent\textit{Flow A\textendash{} Importance Score Pruning.}
\\\indent Each Gaussian’s size $r_{k}$ and opacity $\alpha_{k}$ are combined into a simple importance score
\begin{equation}
s_{k} = \alpha_{k} \cdot r_{k},
\tag{14}
\end{equation}
where $r_{k} = \lVert \boldsymbol{\sigma}_{k} \rVert_{\infty}$ is the maximum standard deviation among its principal axes.
Splats in the lowest $q\%$ of scores are pruned, removing elements that contribute minimally to scene reconstruction.

\FloatBarrier

\begin{center}
  \includegraphics[width=0.85\columnwidth]{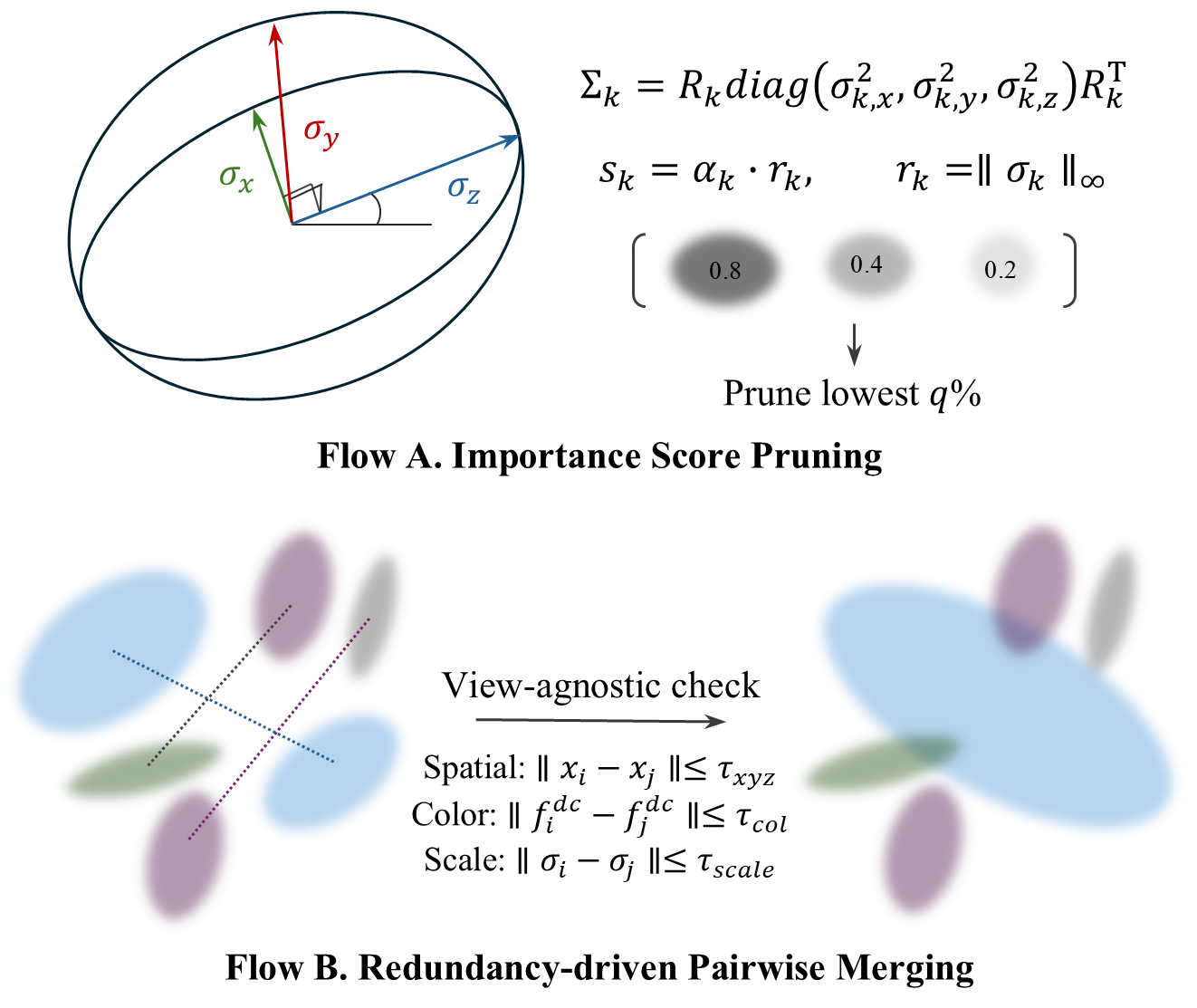}\\[2ex]
  {\footnotesize
   \parbox{\columnwidth}{
     Fig. 5. Stage II: Micro-Refine Phase. Flow A – Rank splats with opacity-and-size and drop the lowest few percent. Flow B – Merge splats that are close in space, similar in base color and size checked in a view-agnostic way and replace each pair with a single splat.
   }
  }
\end{center}

\noindent\textit{Flow B\textendash{} Redundancy-driven pairwise merging.}

Remaining splats are examined in pairs using a view-agnostic proximity test on three attributes: 

\begin{align}
\lVert x_i - x_j \rVert_{2} &\le \tau_{xyz}, \tag{15}\label{eq:crit-xyz}\\
\lVert f_i^{\mathrm{dc}} - f_j^{\mathrm{dc}} \rVert_{2} &\le \tau_{col}, \tag{16}\label{eq:crit-col}\\
\lVert \boldsymbol{\sigma}_i - \boldsymbol{\sigma}_j \rVert_{2} &\le \tau_{\mathrm{scale}}. \tag{17}\label{eq:crit-scale}
\end{align}

\noindent spatial, color, and scale, respectively. If all conditions are met, the pair is merged by averaging each attribute, collapsing duplicates without degrading detail.  
\\\indent Together, these two flows remove the negligible splats and consolidate redundancies, yielding a lean, high-detail model without any separate networks or post-processing.

\section{Results and Evaluation}

\subsection{Implementation}
\indent Our method is implemented in Python using the Pytorch framework ~\cite{ref31}, building on the custom CUDA-based rasterization kernel originally introduced by Kerbel et al. ~\cite{ref6}. We extended these kernels to incorporate our adaptive densification and covariance-based splitting strategies. For interactive viewing and performance measurement, we employed the open-source SIBR library ~\cite{ref32}, which enabled real-time visualization of our rendered scenes. Additionally, we utilized the LPIPS metric ~\cite{ref33} for perceptual evaluations, taking advantage of the publicly available library to quantify subtle differences in image quality.

\begin{center}
  \includegraphics[width=1\columnwidth]{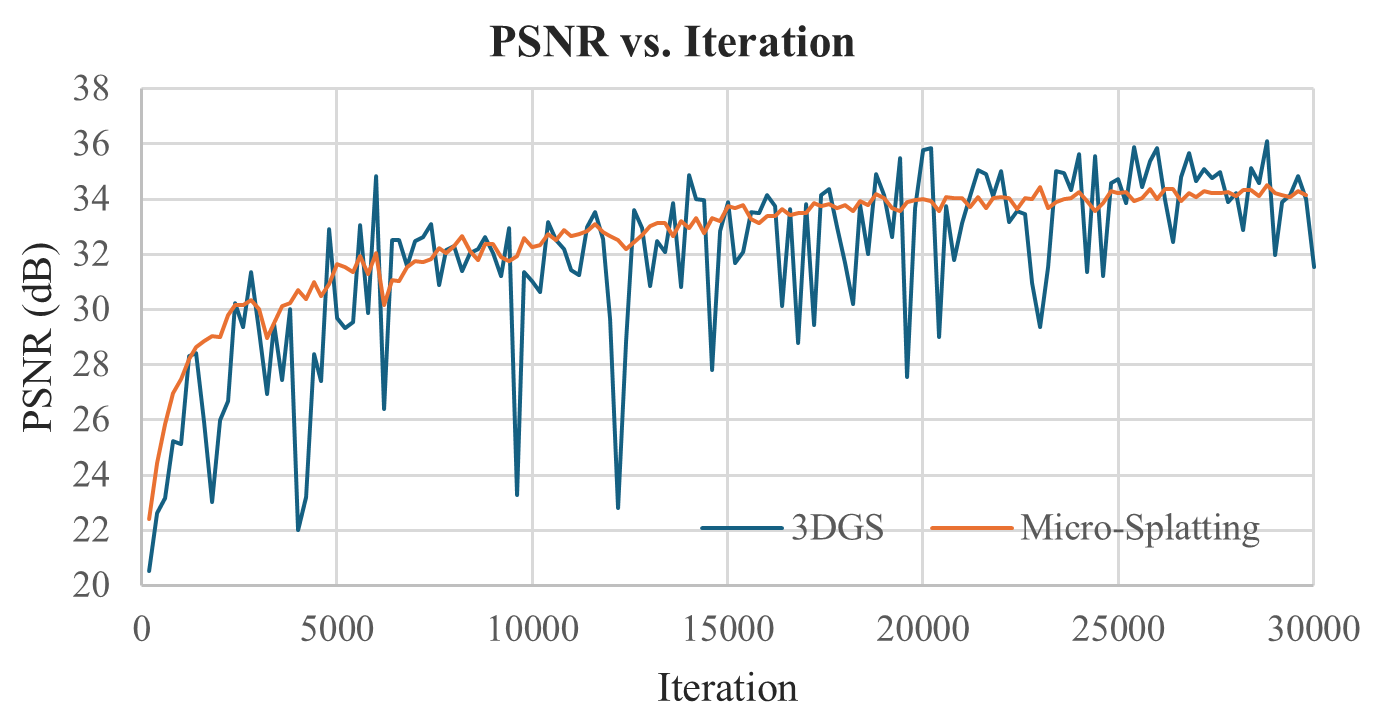}\\[2ex]
  {\footnotesize
   \parbox{\columnwidth}{
     Fig. 6. PSNR vs. Iteration. PSNR progression over 30k training iterations for baseline 3DGS and Micro-Splatting.
   }
  }
\end{center}

\begin{center}
  \includegraphics[width=1\columnwidth]{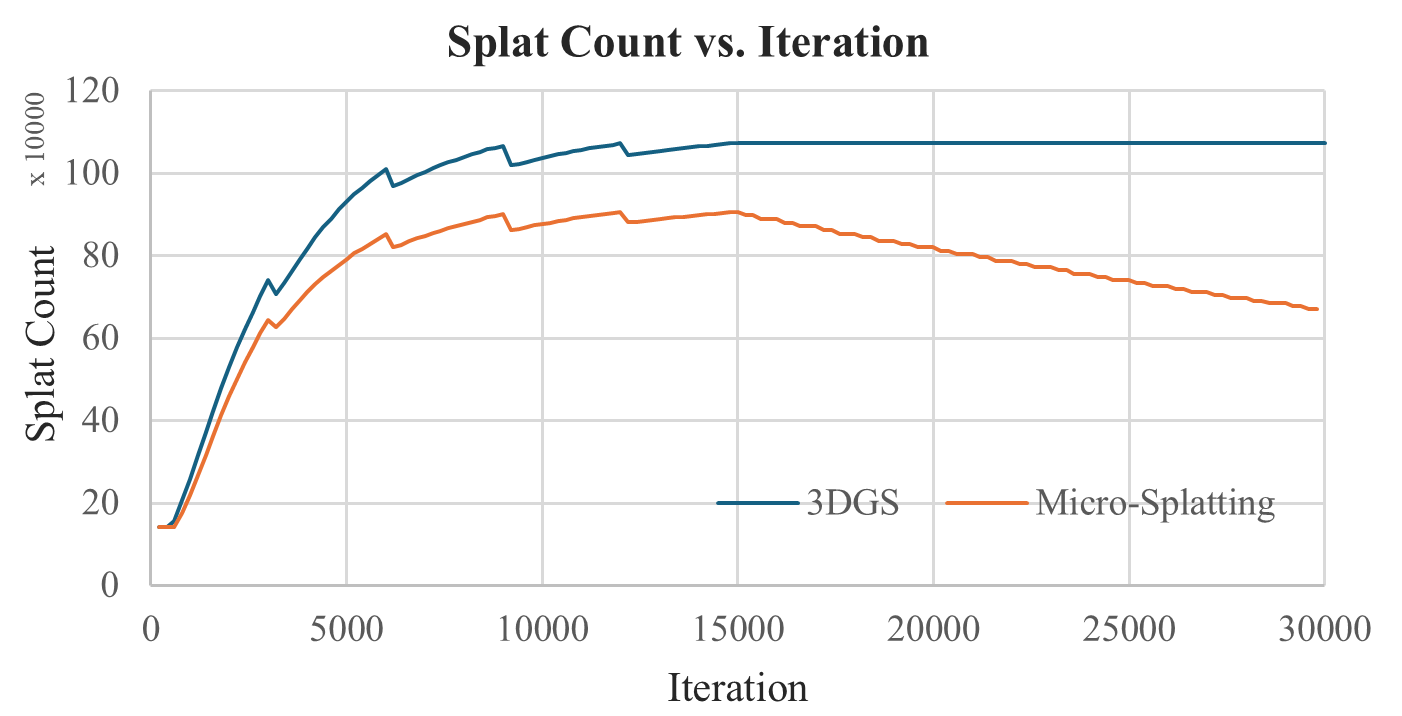}\\[2ex]
  {\footnotesize
   \parbox{\columnwidth}{
     Fig. 7. Splat Count vs. Iteration. Evolution of splat count during training over 30k training iterations for baseline 3DGS and Micro-Splatting.
   }
  }
\end{center}\par
\vspace{1\baselineskip}
\noindent
\subsection{Training Dynamics}
\indent We monitor PSNR on held-out views and the number of active splats throughout training. Micro-Splatting demonstrates a distinct two-phase trajectory: an initial growth phase, spanning approximately 0–15k iterations, in which PSNR increases concurrently with the expansion of the splat set, followed by a micro-refinement phase where the splat count steadily decreases. Notably, Fig. 6 and Fig. 7 clearly show that, even as the number of splats drops after 15k iterations, PSNR continues to rise and eventually stabilizes at a high level. This contrasts sharply with the baseline 3DGS, which exhibits an almost monotonically increasing splat count and a noticeably more irregular PSNR progression at comparable iterations. 
\\\indent The stability and sustained PSNR gain of Micro-Splatting are the result of three integrated design choices. First, targeted splitting densifies only regions with high local reconstruction-gradient magnitudes during Stage I, preventing the disruptive topology changes caused by the broad densify–reset cycles of 3DGS. Second, a trace-based covariance constraint enforces compact, near-isotropic kernels, which mitigates view-dependent instability and avoids transient over- or under-blurring. Third, the inclusion of a mixed L1 and L2 photometric loss supplies error-proportional gradients near convergence, reducing the oscillatory behavior typical of a pure L1 loss. These combined mechanisms enable Micro-Splatting to achieve comparable or superior PSNR with fewer Gaussians and a far steadier optimization trajectory.
\subsection{Splat Distribution in High-Frequency Regions}
\indent To evaluate the behavior of Stage I (Growth), we compare three methods, Pixel-GS ~\cite{ref10}, FreGS ~\cite{ref9}, and Micro-Splatting, all of which share the same objective of concentrating splats in high-frequency regions after 30k iterations. In our approach, a local reconstruction-gradient score $M_{k}$ is computed over each splat’s view-space footprint after every backward pass, and only splats with high $M_{k}$ values which are regions still requiring additional detail are split. Pixel-GS instead drives densification from pixel-aligned residuals, while FreGS employs a global frequency regularizer ~\cite{ref9, ref10}. 
\\\indent {Fig.~8 presents point\mbox{-}cloud views \textit{(top)} and rendered crops \textit{(bottom)} along with per\mbox{-}method splat counts. Although all methods aim to allocate density to textured areas, Micro\mbox{-}Splatting concentrates splats more selectively in genuinely detailed regions such as the flowers in the scene, while keeping low-frequency areas like flat walls sparse \(\big(N_{30\mathrm{k}} \approx 530\mathrm{k}\big)\). In comparison, baseline 3DGS under\mbox{-}densifies the flowers (\(\approx 840\mathrm{k}\)), Pixel\mbox{-}GS substantially overpopulates even the white wall (\(\approx 1.88\mathrm{M}\)), and FreGS similarly increases splat density in low\mbox{-}texture regions (\(\approx 1.17\mathrm{M}\)). These results demonstrate that, despite sharing the same high\mbox{-}frequency targeting goal, Micro\mbox{-}Splatting achieves sharper detail with fewer points by allocating capacity more precisely to truly high\mbox{-}frequency regions.}

\newpage
\setcounter{figure}{7}
\begin{figure*}[t]
  \centering
  \includegraphics[width=\textwidth]{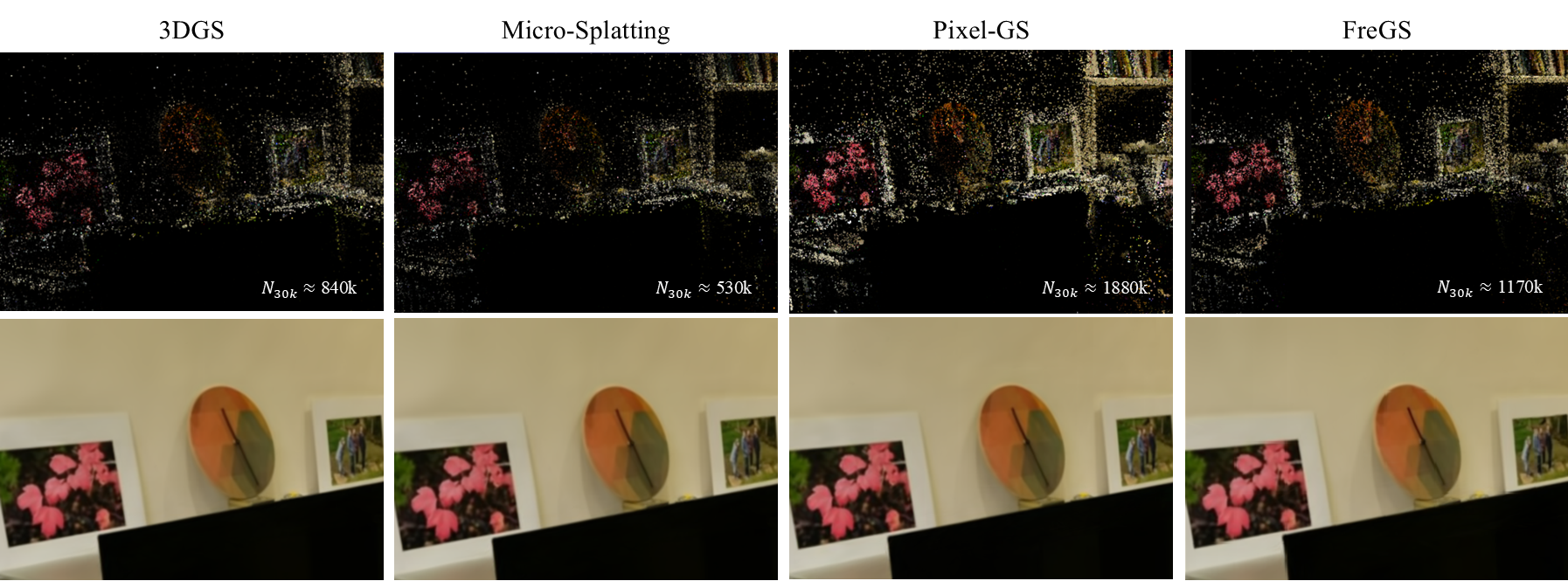}
  \captionsetup{font=footnotesize, name={Fig.}, labelsep=period}
  \caption{Splat distribution vs.\ detail on Mip-NeRF~360~\cite{ref34} room dataset.
  Point-cloud views \textit{(top)} and render crops \textit{(bottom)} at 30k iterations with per-method splat counts ($N_{30\mathrm{k}}$).
  Micro-Splatting concentrates splats on the textured flowers while keeping smooth walls sparse ($N_{30\mathrm{k}}\approx 530\text{k}$).
  Baseline 3DGS~\cite{ref6} under-densifies the flowers ($\approx 840\text{k}$);
  Pixel-GS~\cite{ref10}, though also detail-seeking, heavily overpopulates even the white wall and yields the largest model ($\approx 1.88\text{M}$);
  FreGS~\cite{ref9} similarly increases wall splats and overall count ($\approx 1.17\text{M}$).
  The result is sharper detail with fewer points for Micro-Splatting, and noticeably better allocation of splats to genuinely high-frequency regions.}
  \label{fig:mipnerf-dist}
\end{figure*}

\subsection{Results}
\indent We evaluate Micro-Splatting on five representative Mip-NeRF 360 scenes (bonsai, garden, kitchen, room, counter)~\cite{ref34} and compare against 3DGS ~\cite{ref6}, Pixel-GS ~\cite{ref10}, FreGS ~\cite{ref9}, EDC ~\cite{ref12}, Taming-3DGS ~\cite{ref11}, and Mini-Splatting 2 ~\cite{ref13}, and TrimGS ~\cite{ref16}. All methods are trained for 30k iterations using identical data splits.\par
\vspace{1\baselineskip}
\noindent\textbf{Qualitative Evaluation (Mip-NeRF 360).}
\\\indent Across all scenes, most baselines render foreground objects and nearby structures well but tend to lose detail in distant background regions. In contrast, Micro-Splatting preserves far-field cues more faithfully while maintaining a model size in the same range.  

\begin{itemize}
 \item {Bonsai:} he small figurine visible through the rear window is sharply reconstructed by Micro-Splatting, matching Pixel-GS, while Taming-3DGS and Mini-Splatting 2 blur it away.
 \item {Garden:} The narrow building between the brick wall and the tree remains clearly defined in Micro-Splatting, whereas it is entirely blurred in Taming-3DGS and Mini-Splatting 2, despite only ~50 MB difference in model size and partially lost in FreGS. 
 \item {Kitchen:} The back window emerging from a dark region appears closest to ground truth in Micro-Splatting. 3DGS and Taming-3DGS render it with blur or noticeable color shift, while other baselines smear it out.
 \item {Room: } The TV screen reflection in the far background is sharply reproduced in Micro-Splatting, whereas other methods produce softer, less distinct reflections.
  \item {Counter: } The specular highlight on the distant countertop is preserved in Micro-Splatting and Mini-Splatting 2 but is entirely missing in FreGS and most other baselines.
\end{itemize}
In short, while foreground quality is comparable, Micro-Splatting keeps long-range detail and reflections intact, aligning the ground truth in the far field. \par
\vspace{1\baselineskip}
\noindent\textbf{Quantitative Evaluation (Mip-NeRF 360)}
\\ \textit{{Quantitative quality (Table~1).}} We compare 3DGS~\cite{ref6}, Pixel\mbox{-}GS~\cite{ref10}, FreGS~\cite{ref9}, EDC~\cite{ref12}, Taming~3DGS~\cite{ref11}, Mini\mbox{-}Splatting~2~\cite{ref13}, TrimGS~\cite{ref16}, and our method on five Mip\mbox{-}NeRF~360~\cite{ref34} scenes (\textit{bonsai}, \textit{garden}, \textit{kitchen}, \textit{room}, \textit{counter}), reporting SSIM, $L_{1}$, PSNR, and LPIPS. Micro\mbox{-}Splatting ranks first or second on nearly all metrics, with the highest SSIM and lowest $L_{1}$ on \textit{bonsai}, \textit{kitchen}, and \textit{room}, and competitive PSNR and LPIPS on \textit{garden} and \textit{counter}. Using an $L_{1}+L_{2}+\mathrm{SSIM}$ photometric objective together with trace\mbox{-}regularized kernels yields lower pixel error and stronger structural and perceptual fidelity, helping retain fine detail without introducing high\mbox{-}frequency artifacts.

\medskip

 \noindent\textit{{Efficiency \& size (Table~2).}} For the same scenes and training budget, we report FPS, model size (MB), training time, and splat count ($\times 10^{3}$). Despite concentrating capacity in genuinely high\mbox{-}frequency regions, Micro\mbox{-}Splatting maintains real\mbox{-}time FPS comparable to 3DGS while producing smaller or similar model sizes in most cases. The Stage~II refinement module (importance\mbox{-}score pruning $+$ redundancy\mbox{-}aware merging) cuts splats substantially without quality loss, yielding compact, efficient models. In contrast, several baselines that aggressively densify high\mbox{-}frequency areas (e.g., Pixel\mbox{-}GS, FreGS) often inflate splat count and storage with limited gains. (In both tables, ``Size'' denotes on\mbox{-}disk \texttt{.ply} MB; arrows indicate the preferred direction.)

\clearpage
\begin{figure*}[t]
  \centering
  \includegraphics[width=\textwidth]{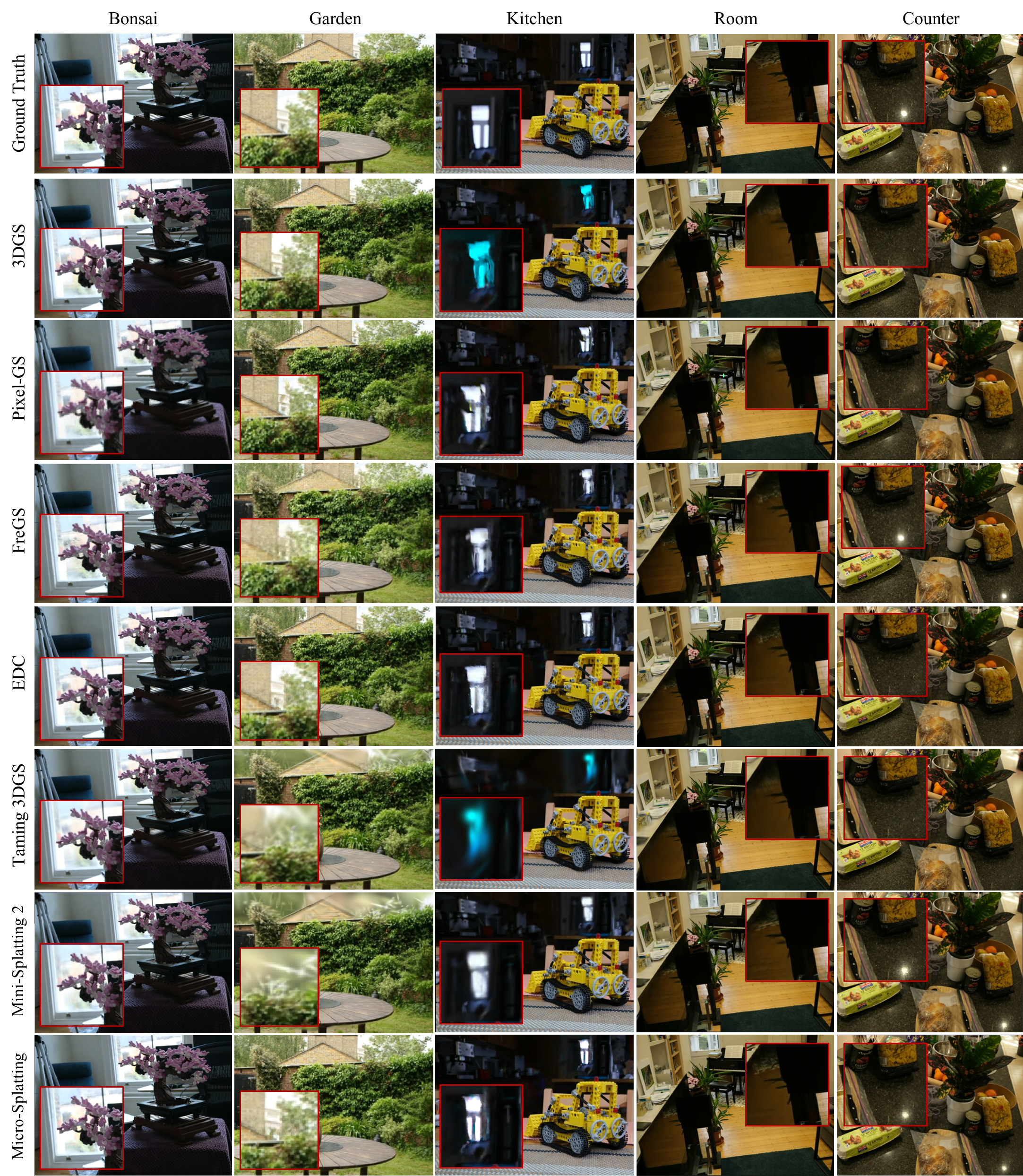}
  \captionsetup{font=footnotesize, name={Fig.}, labelsep=period}
  \caption{Qualitative Comparison on Mip\mbox{-}NeRF~360~\cite{ref34} (\textit{bonsai}, \textit{garden}, \textit{kitchen}, \textit{room}, and \textit{counter}) at 30k iterations. 
  Each triplet shows the ground truth (left), the baseline method (center), and our Micro\mbox{-}Splatting reconstruction (right). 
  Compared to competing methods—3DGS~\cite{ref6}, Pixel\mbox{-}GS~\cite{ref10}, FreGS~\cite{ref9}, EDC~\cite{ref12}, Taming~3DGS~\cite{ref11}, 
  Mini\mbox{-}Splatting~2~\cite{ref13}, TrimGS~\cite{ref16}, and Micro\mbox{-}Splatting—our approach better preserves fine structures and 
  high\mbox{-}frequency details in both foreground and background regions. 
  Notable examples include sharper flower textures (\textit{bonsai}), clearer distant building edges (\textit{garden}), 
  more accurate back\mbox{-}window reconstruction (\textit{kitchen}), well\mbox{-}defined TV reflections (\textit{room}), 
  and intact specular highlights (\textit{counter}). Micro\mbox{-}Splatting achieves this while maintaining compact models and efficient rendering.}
  \label{fig:qual-30k}
\end{figure*}

\clearpage
\begin{table*}[!htb]
  \centering
  \caption{Quantitative Results on Mip\mbox{-}NeRF~360 (30k Iterations)}
  \includegraphics[width=\textwidth]{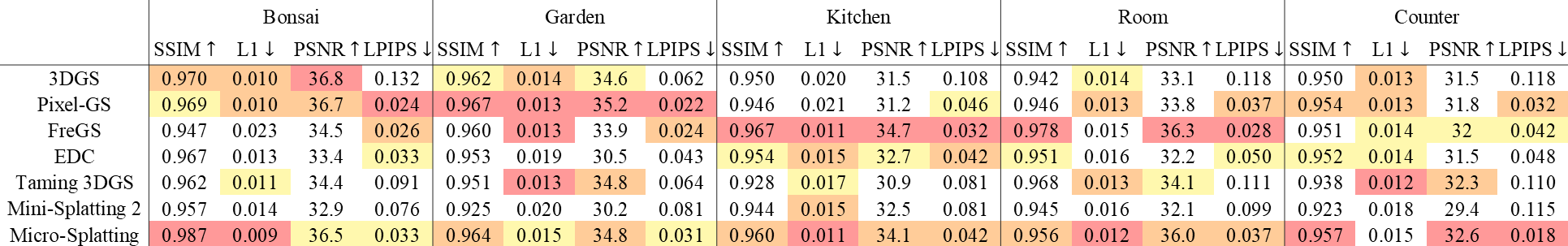}
  \label{tab:mip360-quant-30k}
\end{table*}

\begin{table*}[!t]
  \centering
  \caption{Efficiency and size comparision on Mip\mbox{-}NeRF~360 (30k Iterations)}
  \includegraphics[width=\textwidth]{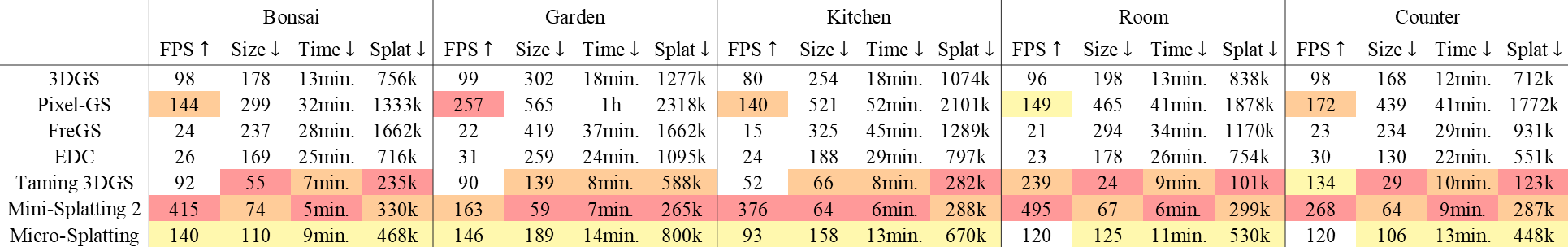}
  \label{tab:mip360-eff-size-30k}
\end{table*}

\subsection{Ablation Study}
\indent We evaluate each component via leave\mbox{-}one\mbox{-}out ablations relative to the full pipeline, using identical splits, iteration budgets, and schedules across all runs. Six variants are tested on eight scenes---five from Mip\mbox{-}NeRF~360~\cite{ref34} (\textit{kitchen}, \textit{bonsai}, \textit{garden}, \textit{counter}, \textit{room}) and three from Tanks \& Temples~\cite{ref35} (\textit{Ignatius}, \textit{Church}, \textit{Truck}): (a) Full Micro\mbox{-}Splatting, (b)~-- Covariance regularization, (c)~-- High\mbox{-}frequency focus, (d)~-- Importance\mbox{-}score pruning, (e)~-- Redundancy\mbox{-}driven merging, (f)~-- Isotropy enforcement. We report PSNR, SSIM, LPIPS, FPS, model size (MB; on\mbox{-}disk \texttt{.ply}), and splat count ($\times 10^{3}$). Table~3 lists the average results over all eight scenes, with detailed per\mbox{-}scene metrics provided in Appendix~I. Table~4 provides a per\mbox{-}scene breakdown for the \textit{kitchen} dataset, and Table~5 shows the radius\mbox{-}bin distribution (Bin~0~--~Bin~9), which partitions the mean 2D footprint radius into ten equal\mbox{-}width intervals at the training resolution.

\begin{description}[font=\itshape, labelsep=0.6em, leftmargin=2.6em, style=nextline, itemsep=0.3em]
  \item[(a)] Full Micro-Splatting:
  \\\indent Delivers the best overall balance, with top SSIM/PSNR, low LPIPS, compact size, and high FPS. Coverage is dense only in textured areas, with radius bins favoring small kernels and smooth tapering for fine detail without overgrowth.

  \item[(b)] No Covariance Regularization: 
\\\indent Without the trace cap, kernels over-expand or elongate, causing directional blur and higher pixel error. The radius-bin histogram skews toward extremes, confirming unstable growth.

  \item[(c)] No High-Frequency Focus:
\\\indent Disabling gradient-driven splitting leads to sparse coverage and missed fine details, producing the largest quality drop among Stage I changes and collapsing small-radius bin counts.

  \item[(d)] No Importance-Score Pruning:
\\\indent Quality remains similar, but splat count, size, and training time rise sharply. Extra splats clutter already accurate regions, showing that pruning removes low-impact points efficiently.

  \item[(e)] No Redundancy-Driven Merging:
\\\indent Near-duplicate splats remain, increasing memory use and reducing FPS. Dense areas show mild over-smoothing where merging would yield cleaner results.

  \item[(f)] No Isotropy Enforcement:
\\\indent Anisotropic kernels introduce smear and color bleeding along high-contrast edges, yielding poor LPIPS/PSNR despite moderate splat counts.
\end{description}

\indent The ablation study shows that Stage I components (Mk-based splitting, covariance trace cap, and isotropy enforcement) are key to producing stable, fine detail, while Stage II components (importance-score pruning and redundancy merging) maintain compactness without sacrificing fidelity. Together, these yield the best scores and cleanest reconstructions in Fig. 10, while controlling memory usage and runtime.

\begin{table}[H]
  \centering
  \caption{Quantitative ablation results averaged over 8 scenes   \\(5 Mip\mbox{-}NeRF~360; 3 Tanks \& Temples)}
  \includegraphics[width=0.95\columnwidth]{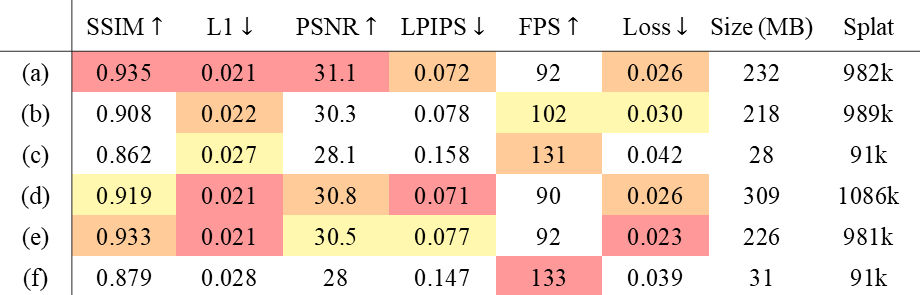}
  \label{tab:abl-avg8}
\end{table}
\clearpage

\begin{table*}[!t]
  \centering
  \begingroup \captionsetup{font=footnotesize,labelfont=bf,labelsep=period,
                name={Table},aboveskip=0pt,belowskip=6pt}
  \begin{minipage}[t]{0.49\textwidth}
    \centering
    \captionof{table}{Quantitative ablation results on Mip\mbox{-}NeRF~360 \\(Kitchen scene, 30k iterations)}
    \includegraphics[width=\linewidth]{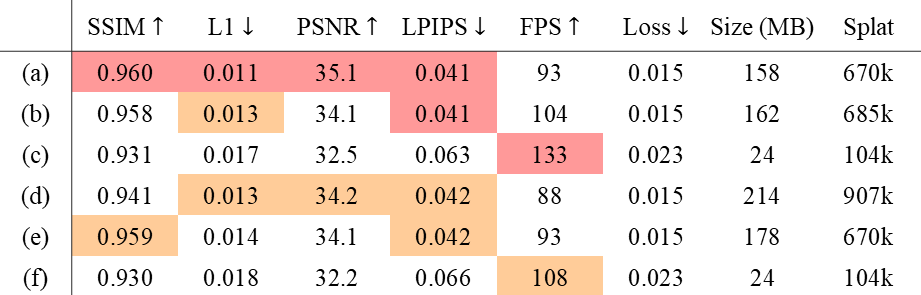}
    \label{tab:abl-kitchen-quant}
  \end{minipage}
  \hfill
  \begin{minipage}[t]{0.49\textwidth}
    \centering
    \captionof{table}{Splat Cardinality of ablation study on Mip\mbox{-}NeRF~360 \\(Kitchen scene, 30k iterations)}
    \includegraphics[width=\linewidth]{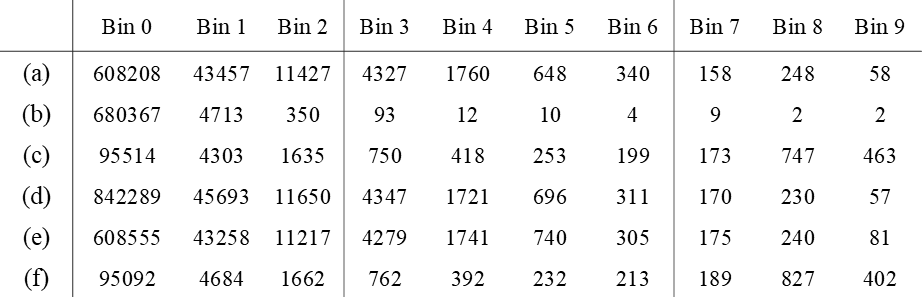}
    \label{tab:abl-kitchen-radius}
  \end{minipage}

  \endgroup
\end{table*}

\begin{figure*}[!t]
  \centering
  \includegraphics[width=\textwidth]{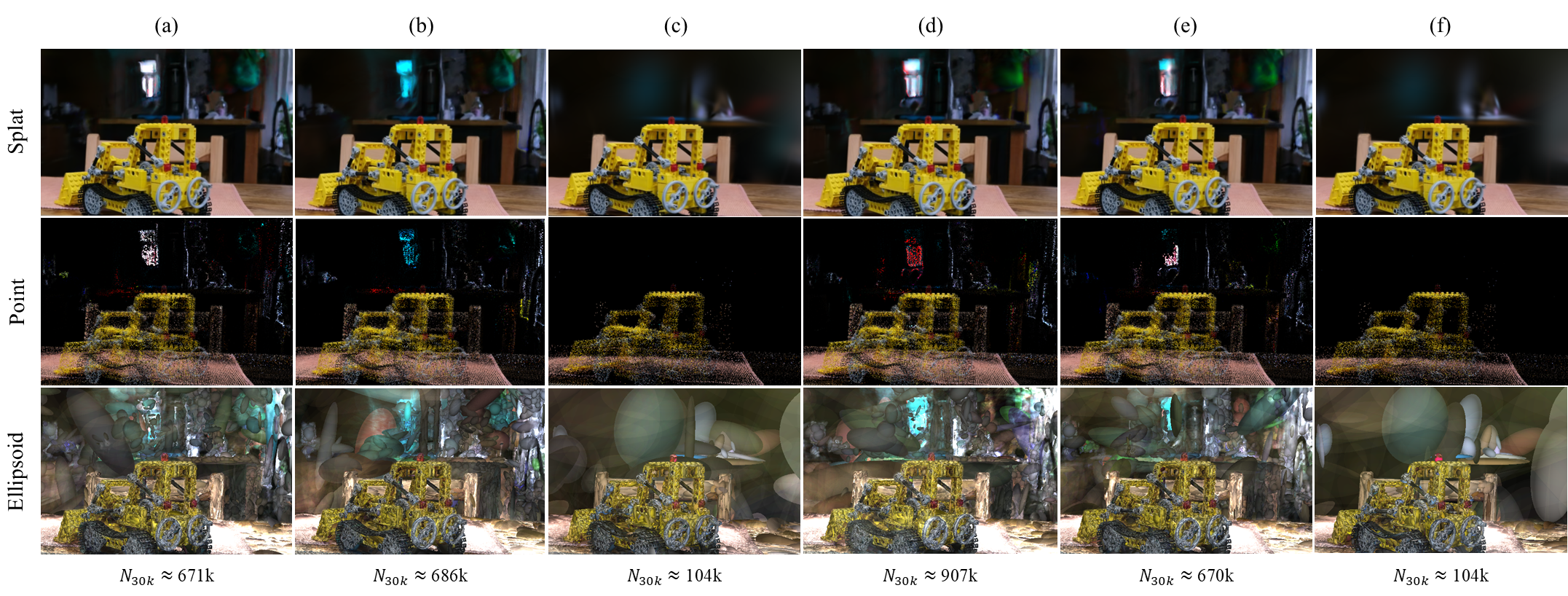}
  \captionsetup{font=footnotesize, name={Fig.}, labelsep=period}
  \caption{Qualitative ablation results on Mip\mbox{-}NeRF~360~\cite{ref34} (Kitchen scene, 30k iterations).
  Comparison of full Micro\mbox{-}Splatting \textit{(a)} and its ablated variants \textit{(b--f)} at 30k iterations, shown in splat render,
  point\mbox{-}cloud, and ellipsoid views. The full model maintains sharp details with dense, well\mbox{-}placed splats, while ablations exhibit effects such as
  blur, under\mbox{-}densification, kernel over\mbox{-}expansion, or excessive redundancy, as reflected in the final splat counts $N_{30\mathrm{k}}$.}
  \label{fig:ablation-qual}
\end{figure*}

\section{Discussion and Conclusion}
\noindent \textit{Discussion}. Our experiments consistently demonstrate that Micro-Splatting enhances fine-detail reconstruction while maintaining compact models. This is supported by three key findings. First, the training dynamics exhibits a clear two-phase progression: a targeted Stage I growth that selectively increases splats in high-frequency regions, followed by a Stage II micro-refinement that steadily reduces splat count even as PSNR continues to rise. Second, splat-distribution visualizations confirm that the view-space, gradient-guided split rule concentrates density only where texture is present, avoiding oversampling in smooth, low-detail areas. Third, the ablation study validates the role of each component: trace-capped covariance and isotropy enforcement stabilize kernel shapes and spherical harmonics fitting; the local $M_{k}$ trigger recovers high-frequency structures; and importance-score pruning combined with redundancy-aware merging reduces splat count without perceptible quality loss. Compared to 3DGS, Micro-Splatting yields smoother PSNR curves, attributable to the trace cap and mixed $L_{1} + L_{2}$ photometric loss. Importantly, these gains are achieved through lightweight computations without auxiliary networks or costly post-processing, keeping training time on par with baselines and preserving real-time rendering performance.\par
\vspace{1\baselineskip}
\noindent \textit{Conclusion}. Micro-Splatting presents a unified, end-to-end framework that first concentrates growth in genuinely high-frequency regions and then refines to a compact, efficient set of Gaussians. By combining trace-based covariance control with isotropy enforcement, view-space gradient-guided splitting, and a lightweight refinement stage, the method achieves sharper textures, reduced artifacts, and significantly smaller model sizes compared to strong baselines, all while sustaining real-time rendering performance.

\section{Acknowledgment}

\noindent This work was supported by the National Research Foundation of Korea (NRF) grant funded by the Korea government (MSIT) (Grant No. RS-2022-NR067080 and RS-2025-05515607).

\section{Author Contribution}
\noindent
Jee Won Lee: Data Curation, Methodology, Validation, Visualization, Writing--Original draft, Software.\\
Hansol Lim: Data Curation, Visualization, Validation.\\
Sooyeun Yang: Data Curation, Formal Analysis, Writing--Original draft.\\
Jongseong Brad Choi: Conceptualization, Methodology, Validation, Formal Analysis, Writing--Review \& Editing.

\clearpage
\onecolumn        
\counterwithin{figure}{section}
\counterwithin{table}{section}
\renewcommand{\thesection}{\Roman{section}}
\numberwithin{equation}{section}

\section*{APPENDIX I.\ Per-Scene Ablation Results}
\addcontentsline{toc}{section}{APPENDIX I. Per-Scene Ablation Results (8 Scenes)}

\vspace{1\baselineskip}
\noindent This appendix reports concise per-scene results for the six variants defined in Section V. E. — (a) Full Micro-Splatting, (b) – Covariance regularization, (c) – High-frequency focus, (d) – Importance-score pruning, (e) – Redundancy-driven merging, (f) – Isotropy enforcement—under identical splits, iteration budgets, and schedules. We list PSNR/SSIM/LPIPS, FPS, model size (MB), splat count (×10³), and radius-bin distributions. Section V. E. summarizes averages over all 8 scenes (Table 4); a full per-scene example for kitchen appears in Table 4 and Fig. 10.

\vspace{1\baselineskip}
\noindent\textbf{Mip-NeRF 360 Dataset.}
Mip-NeRF 360 ~\cite{ref34} is a widely used multi-view dataset with 360° camera coverage and unbounded scene extent. It contains challenging indoor scenes with glossy surfaces, thin structures, high-frequency textures, and large low-texture regions—conditions that expose aliasing and over-smoothing in 3DGS. We evaluate \textit{kitchen}, \textit{bonsai}, \textit{garden}, \textit{counter}, \textit{room}, using identical splits and training budgets across all ablations.

\vspace{1\baselineskip}
{%
\renewcommand{\thefigure}{A\arabic{figure}}
\setcounter{figure}{0}
\begin{figure}[!htbp]
  \centering
  \includegraphics[width=\columnwidth]{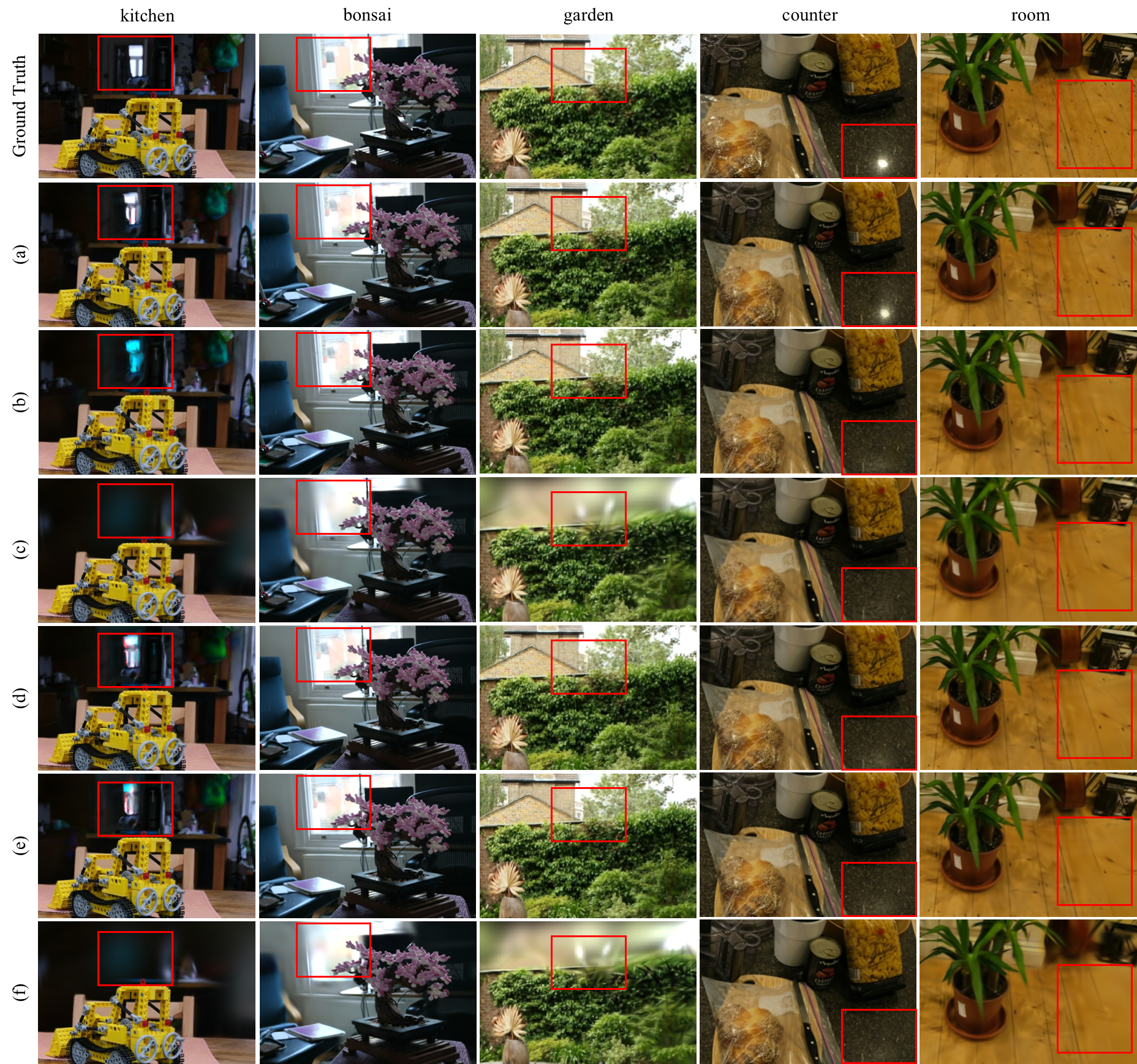}
  \caption{Qualitative results of the ablation study on Mip\mbox{-}NeRF~360~\cite{ref34}:
  Per\mbox{-}scene ablation (\textit{kitchen}, \textit{bonsai}, \textit{garden}, \textit{counter}, \textit{room}) over 30k iterations.
  Qualitative comparisons for variants \textit{(a)} Full, \textit{(b)} -- Covariance reg., \textit{(c)} -- High\mbox{-}freq focus,
  \textit{(d)} -- Pruning, \textit{(e)} -- Merging, \textit{(f)} -- Hard\mbox{-}isotropy. 
  Red boxes mark challenging high\mbox{-}frequency regions; Full Micro\mbox{-}Splatting preserves edges and textures while variants (b,c)
  either blur detail or misallocate capacity.}
  \label{fig:A1}
\end{figure}
}%
\makeatletter

\setlength{\@fptop}{0pt}         
\setlength{\@fpsep}{8pt}          
\setlength{\@fpbot}{0pt plus 1fil} 
\setlength{\@dblfptop}{0pt}       
\setlength{\@dblfpsep}{8pt}
\setlength{\@dblfpbot}{0pt plus 1fil}
\makeatother

\begingroup
  \counterwithout{table}{section}
  \renewcommand{\thetable}{A\arabic{table}}
  \setcounter{table}{0}

  \begin{table}[p]
    \centering
    \captionsetup{font=footnotesize, labelfont=bf, name={Table}, labelsep=period}
    \caption{Quantitative Results and Splat Cardinality of Ablation Study on Mip\mbox{-}NeRF~360.}
    \vspace{0.5ex}
    \includegraphics[width=\textwidth,height=0.92\textheight,keepaspectratio]{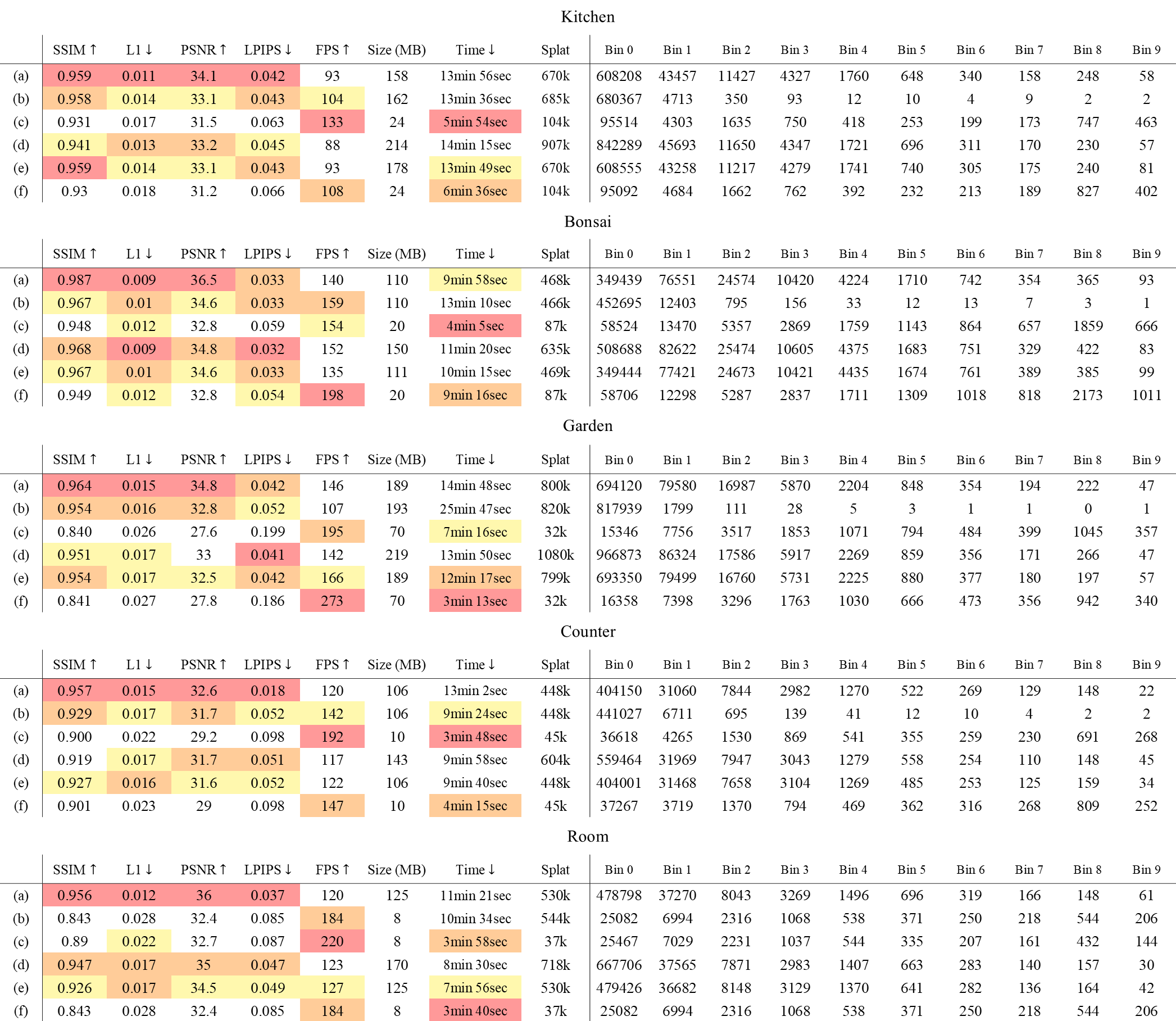}
    \label{tab:A1}
  \end{table}
\endgroup

\FloatBarrier

\vspace{1\baselineskip}
\noindent\textbf{Tanks \& Temples Dataset.}
Tanks \& Temples ~\cite{ref35} is a large-scale multi-view benchmark with wide baselines, complex geometry, and challenging lighting. We evaluate \textit{Ignatius}, \textit{Church}, \textit{Truck}, which stress high-frequency stone relief, repetitive masonry, and thin metallic edges/reflections. All ablations use identical splits and training budgets as in Section V. E.

\vspace{1\baselineskip}
{%
\renewcommand{\thefigure}{A\arabic{figure}}
\setcounter{figure}{1}
\begin{figure}[!htbp]
  \centering
  \includegraphics[width=\columnwidth]{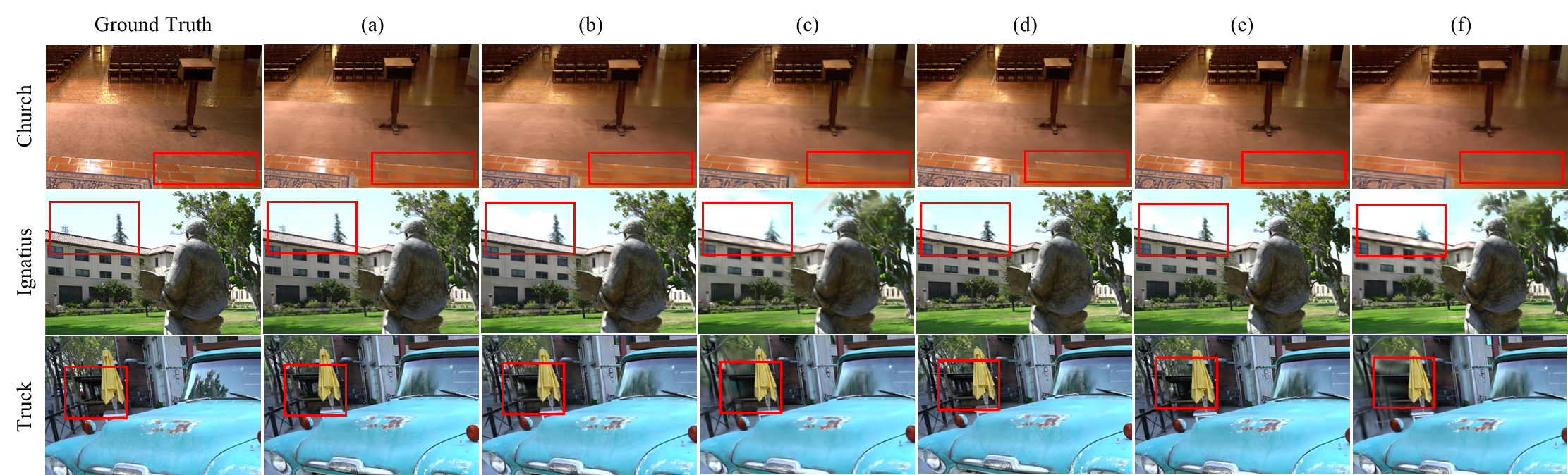} 
  \caption{Qualitative results of the ablation study on Mip\mbox{-}NeRF~360~\cite{ref34}:
  Per\mbox{-}scene ablation (\textit{kitchen}, \textit{bonsai}, \textit{garden}, \textit{counter}, \textit{room}) over 30k iterations.
  Qualitative comparisons for variants \textit{(a)} Full, \textit{(b)} -- Covariance reg., \textit{(c)} -- High\mbox{-}freq focus,
  \textit{(d)} -- Pruning, \textit{(e)} -- Merging, \textit{(f)} -- Hard\mbox{-}isotropy. 
  Red boxes mark challenging high\mbox{-}frequency regions; Full Micro\mbox{-}Splatting preserves edges and textures while variants (b,c)
  either blur detail or misallocate capacity.}
  \label{fig:A1}
\end{figure}
}%

\FloatBarrier

\begingroup
  \counterwithout{table}{section}
  \renewcommand{\thetable}{A\arabic{table}}
  \setcounter{table}{1}

  \begin{table}[!htbp]
    \centering
    \captionsetup{font=footnotesize, labelfont=bf, name=Table, labelsep=period}
    \caption{Quantitative Results and Splat Cardinality of Ablation Study on Mip\mbox{-}NeRF~360.}
    \vspace{0.5ex}
    \includegraphics[width=\columnwidth]{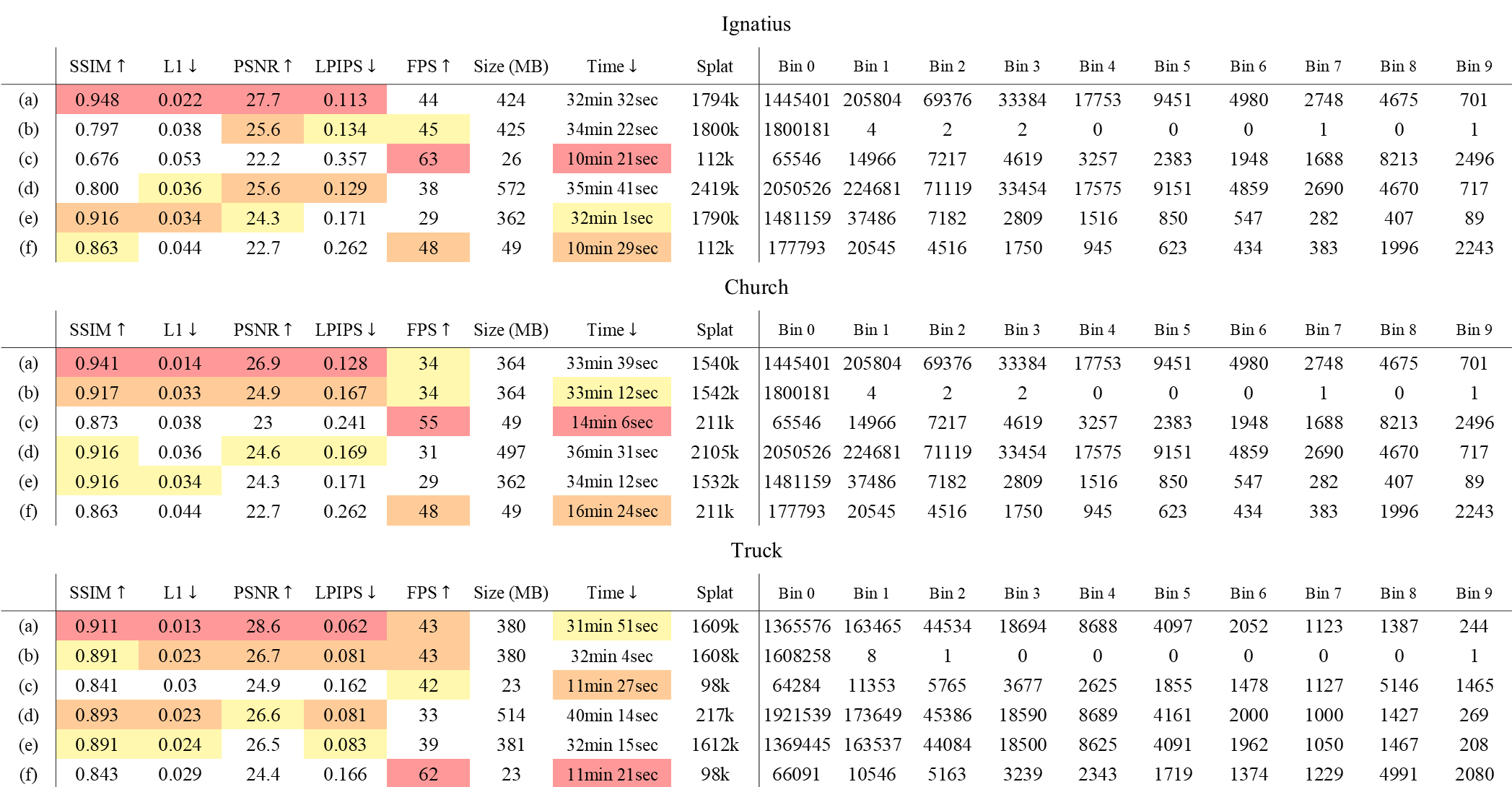}
    \label{tab:A2}
  \end{table}
\endgroup

\end{document}